# Finite-size and composition-driven topological phase transition in $(Bi_{1-x}In_x)_2Se_3$ thin films

Maryam Salehi[‡, §], Hassan Shapourian[#, §], Nikesh Koirala[†], Matthew J. Brahlek[†, ‖], Jisoo Moon[†] and Seongshik Oh[†, *]

[‡]Department of Materials Science and Engineering, Rutgers, The State University of New Jersey, Piscataway, New Jersey 08854, U.S.A.

[#]Department of Physics, University of Illinois at Urbana-Champaign, Urbana, Illinois 61801, USA

[†]Department of Physics and Astronomy, Rutgers, The State University of New Jersey, Piscataway, New Jersey 08854, U.S.A.

*Correspondence should be addressed to **ohsean@physics.rutgers.edu**, Fax: 732-445-4343

ABSTRACT: In a topological insulator (TI), if its spin-orbit coupling (SOC) strength is gradually reduced, the TI eventually transforms into a trivial insulator beyond a critical point of SOC, at which point the bulk gap closes: this is the standard description of the topological phase transition (TPT). However, this description of TPT, driven solely by the SOC (or something equivalent) and followed by closing and reopening of the bulk band gap, is valid only for infinite-size samples, and little is known how TPT occurs for finite-size samples. Here, using both systematic transport measurements on interface-engineered $(Bi_{1-x}In_x)_2Se_3$ thin films and theoretical simulations (with animations in Supporting Information) we show that description of TPT in finite-size samples needs to be substantially modified from the conventional picture of TPT due to surface-state hybridization and bulk confinement effects. We also show that the finite-size TPT is composed of two separate transitions, topological-normal transition (TNT) and metal-insulator transition (MIT) by providing a detailed phase diagram in the two-dimensional phase space of sample size and SOC strength.

Key Words: Topological insulator, topological phase transition, metal to insulator transition, molecular beam epitaxy, thin films

Topological insulators (TIs) are a class of electronic materials with insulating bulk and strong spin-orbit coupling (SOC) which supports gapless Dirac topological surface states (TSS) at their boundary with non-topological (trivial) band insulators[1-6]. If we start from a bulk sample of archetypal 3D-TI $Bi_2Se_3$ [7-11] and move toward its trivial insulator counterpart $In_2Se_3$ (with the same $D_{3d}^{5}$ crystal structure[12]) by varying indium (In) concentration in solid solution $(Bi_{1-x}In_x)_2Se_3$, there has to be a critical x at which the bulk gap closes and topological phase transition (TPT) from TI to non-TI occurs and beyond which the bulk gap reopens and bulk bands de-invert to their atomic limit ordering[13, 14]. Upon adding In and as the bulk gap becomes smaller and the penetration depth of topological surface state (TSS) keeps growing until it diverges at the critical point where the surface states eventually disappear by merging into the bulk states. Therefore, as the sample is made thinner, the diverging thickness of the top and bottom surface states becomes comparable with the film thickness, and hence TSSs hybridize and open a gap at the Dirac point converting the system into a trivial insulator before In concentration reaches the bulk critical point. Furthermore, even the bulk states are affected by the quantum confinement effect. However, such finite-size-driven TPT, which is quite different from the conventional TPT process envisioned from infinite-size samples, has been neither observed experimentally due to lack of proper materials nor studied theoretically.

Ideally, TIs must be metallic on the topological side of the TPT (we name it topological-metallic) and insulating on the non-TI side (we name it normal-insulating). However, the system always remained metallic on both sides of the transition due to surface Fermi level being located in the bulk conduction band in previous studies[14]. Consequently, such a TPT directly associated with metal-insulator transition (MIT) has never been observed and at best, only weak signatures of increased scattering around the transition point have been observed[15]. Here, by utilizing a new generation of bulk insulating $(Bi_{1-x}In_x)_2Se_3$ thin films, we investigate, for the first time, finite-size driven TPT via transport measurements and theoretical simulations.

Thin films with thicknesses of 5, 6, 7, 8, 10, 12, 20, and 40 QL (1 QL ≈ 1 nm) $(Bi_{1-x}In_x)_2Se_3$ (0 < x < 0.1) were grown on insulating $In_2Se_3/(Bi_{0.5}In_{0.5})_2Se_3$ buffer layer[16] on sapphire substrate (key features of growth and transport properties are described in the Supporting Information Section IA and IB; details in Ref.16). In this new-generation of $(Bi_{1-x}In_x)_2Se_3$ films, the surface Fermi level ($E_F$) is already below the bottom of the conduction band[16, 17] with low sheet carrier densities ($n_{sheet} \approx 1 - 3 \times 10^{12}$ cm$^{-2}$ ) for the entire thickness range of 5 to 60 QL. With *in situ* deposition of 50 nm electron-depleting $MoO_3$-capping[18], the surface $E_F$ can be pushed even closer to the Dirac point with the carrier density reduced almost by half (details in Ref. 16). $MoO_3$-capping can also function as a protection-layer against aging in air for these low-carrier-density films, especially if it is combined with selenium capping[19].

Figures 1a-h show the sheet resistance as a function of temperature for each thickness and with different In concentration. The 5 QL-thick sample without In shows insulating behavior (Figure 1a) due to the presence of the hybridization gap at the Dirac point and the surface $E_F$ being close to (or inside) this surface gap. Also the 6 QL-thick sample without any In (Figure 1b), which is on the verge of surface gap opening, is weakly insulating. These observations are in contrast with previous-generation samples[14, 20-22] with their surface $E_F$ above the bulk conduction band, which exhibit metallic temperature dependence even for films as thin as 3 QL (Supporting Information Section IB). Moreover, Figure 1 shows that the thickness, where metal-insulator transition (MIT) occurs, grows toward 7 QL, 8 QL and 10 QL as the In concentration grows to 3%, 4%, and 5%, respectively. This observation suggests that the thickness of the surface states and thus the critical thickness, where the surface hybridization gap starts to form, grows as the strength of SOC decreases. Further, the Ioffe-Regel criterion can be used to quantitatively identify the MIT point. Following this criterion, a material becomes metal if $k_F l > 1$ and insulator if $k_F l \leq 1$, where 2D Fermi vector $k_F = (2\pi n_{sheet})^{1/2}$ and mean-free path $l = (\hbar\mu/e)(2\pi n_{\text{sheet}})^{1/2}$ [23, 24]. Figure 1i shows the $k_F l$ as a function of In concentration for each thickness of $(Bi_{1-x}In_x)_2Se_3$ sample. The intersection of the dashed line at $k_F l = 1$ and each curve roughly determines the critical In concentration for MIT. Additionally, as the film goes through MIT, we observed that the number of conducting channel ($\tilde{A}$) in weak anti-localization analysis collapses to zero (See Supporting Information Section IB).

Although the thickness-dependent MIT is the direct manifestation of TPT, this measurement alone does not provide the full picture of the transition. In Figure 2a, the evolution of band structure as a function of SOC for different thicknesses is computed using a simple four-band massive Dirac model[25] (details in Supporting Information Section II), which further provides many of the important insights and details that are not well captured by the experiment regarding the general issues of TPT. In addition, orbital content is embedded as color in the band structure where blue is even parity (Bi-like p orbital), red is odd parity (Se-like p orbital), and yellow is 50-50 mixture of Bi and Se (going from even to odd parity) representing the topological surface states (see two other color schemes corresponding to the spin and confinement in Supporting Information Section II along with six animations simulating the TPT in thin and thick regimes explained in Supporting Information Section III). This color scheme reveals that in the TI phase there is a band inversion near the Γ point (i.e. the conduction band has more Se-characteristic, whereas the valence band has more Bi- characteristic), while the bands away from the Γ point are not affected by the band inversion.

One notable feature in the simulation (Figure 2a, Figure 3b and Supporting Information Figure S5) is that the sample thickness makes a significant impact on the TPT process both qualitatively and quantitatively. In the thin regime the bulk gap keeps decreasing upon weakening SOC (by adding In), which causes TSSs gradually hybridize (at the Dirac point) and start to merge into the bulk states (this is clear for 10 QL sample where yellow turns to blue/red right after gap opening in Figure 2a). In this process, the bulk gap never completely closes (Figure 3b) and this is evident from the fact that the minimum energy of an electron always remains finite due to the spatial confinement in thin films. However, in the thick regime TSSs remain gapless until the critical transition point where the bulk conduction and valance band meet at the Γ point and TSSs merge into the bulk states. After this point, the bulk gap increases and the bulk bands de-invert (see Figure 2a, third panel).

The phase diagram in Figure 2b summarizes the main result of this paper. It consists of three different regions: topological-metallic (TM), normal-metallic (NM), and normal-insulating (NI). This phase diagram focuses on the thickness ≤ 20 QL and In concentration < 10%, where the finite-size effect on the topological phase transition is most clear. The black solid curve defines the boundary of topological-to-normal transition (TNT), which can qualitatively be thought to occur when the sample thickness equals twice the penetration depth of the surface states. Although the TNT line cannot be detected with our transport measurement, it was identified previously with terahertz spectroscopy as the scattering rate jumps through this line [15]; the line separating hybrized TI and 3D TI in Ref. 15 is equivalent to the TNT line in the present phase diagram. The TNT line starts from 6 QL for 0% In according to previous ARPES studies on pure $Bi_2Se_3$ films[15, 25], and approaches asymptotically the bulk critical value, which is not yet known precisely due to lack of reliable data for bulk samples and are thus assumed to be 7% as the best guess from other measurements on finite-thickness samples[14, 15]. It is also notable that in contrast to previous thoughts, there is not a phase with hybridized surface states and topologically non-trivial bulk states; once the surface states hybridize and open a gap, the system is no longer topological and should be considered normal. We confirm this behavior by directly computing the band inversion for finite-size samples (see Supporting Information Section II), and thus, the earlier phase diagram in Ref. 15 proposing such a phase needs to be modified. Further, even after the samples become normal (non-topological), they do not immediately become insulating. In other words, TNT does not coincide with MIT, and their extent of separation is determined by the surface $E_F$ as shown in Figures 2c and d for two extreme cases. In the ideal case (Figure 2c), where $E_F$ is exactly at the Dirac point, as soon as a gap is formed at the Dirac point, the sample becomes insulating: in other words, TNT coincides with MIT. On the other hand, if $E_F$ is high and resides in the conduction band (Figure 2d), the sample remains fully metallic even after the topological surface band completely disintegrates away due to the high $E_F$; this has been the case for all previous samples used for TPT studies[13-15, 26, 27].

Now if the surface $E_F$ is close to, but not exactly at the Dirac point, TNT and MIT are separated from each other in proportion to the $E_F$ as shown in Figure 2b. More specifically, as In concentration grows or sample thickness decreases, a gap develops at the Dirac point of the surface state going through TNT, but the film still remains metallic because $E_F$ is not at or near the gap. Then as the gap further grows, the sample eventually becomes insulating going through MIT. These two transitions, TNT and MIT, together complete the TPT process. Another important factor here is that it is not the Dirac gap but the mobility edge that determines the separation between TNT and MIT. The presence of a mobility edge shrinks the gap between TNT and MIT and thus, makes the samples behave close to the ideal case (Figure 2c). Even if we do not know the exact location of the mobility edge, a simple analytic approximation for the position of the mobility edge reproduces the MIT boundary (Figure 3a; for more details see Figure S6 in Supporting Information), qualitatively consistent with the experimental observation (Figure 2b).

For further clarification Figure 3b shows that in thick samples, the critical point is where the bulk band gap closes, and exactly at this point, the bulk-inversion starts to appear/disappear and the surface band also starts to appear/disappear. In contrast, in finite-thickness samples, the bulk band gap never closes completely due to the confinement effect; moreover, the point where bulk band gap reaches its minimum does not coincide with the point where the hybridization gap starts to develop (for more details see Figure S5 of Supporting Information).

In conclusion, by combining the transport studies of low-Fermi-level $(Bi_{1-x}In_x)_2Se_3$ thin films and theoretical simulations, we provide the first in-depth study of the finite-size effect on the TPT process. The TPT process is composed of TNT, where a gap is formed at the Dirac point, and MIT, where the entire system becomes insulating. While the TNT line is determined by intrinsic material parameters such as sample thickness and In concentration, the MIT line is determined by $E_F$ relative to the mobility edge, both of which depend strongly on sample qualities. As the samples become thin, the critical point for TNT becomes less well-defined and shifts toward lower In concentration, and the bulk band gap never closes near the critical point due to the confinement effect. These findings have important messages for any topological 3D systems including Dirac and Weyl semimetals, which eventually need to be implemented in thin films for device applications.

Experimental Method: Films were grown on 10 mm × 10 mm $Al_2O_3$(0001) substrates using custom built SVTA MOS-V-2 MBE system with base pressure of ~$10^{-10}$ Torr. Substrates were cleaned *ex situ* by 5 minutes exposure to UV-generated ozone and *in situ* by heating to 800 ºC in oxygen pressure of 1 × $10^{-6}$ Torr for 15 minutes. 99.999% pure elemental Bi, In and Se sources were thermally evaporated using Knudsen cells for film growth. Source fluxes were calibrated *in situ* by quartz crystal microbalance (QCM) and *ex situ* by Rutherford backscattering spectroscopy (RBS). The ratio of Se flux to combined Bi and In flux was maintained at above 10:1 during the growth.

The Hall effect and magneto-resistance were measured in magnetic field up to 0.6 T at 6 K along with longitudinal resistance as a function of temperature for all samples in the van der Pauw (VDP) geometry using pressed indium wires as contact leads. Sheet carrier density ($n_{sheet}$) was extracted from $n_{sheet} = (e\, dR_{Hall}/dB)^{-1}$ where $e$ is the electronic charge and $dR_{Hall}/dB$ is the slope of the Hall resistance as a function of magnetic field $B$. The Mobility μ was calculated using $\mu = (e\, R_{sheet}\, n_{sheet})^{-1}$, where zero-field sheet resistance $R_{sheet} = R_{XX}(B = 0)\, \pi/\ln(2)$ for the VDP geometry.

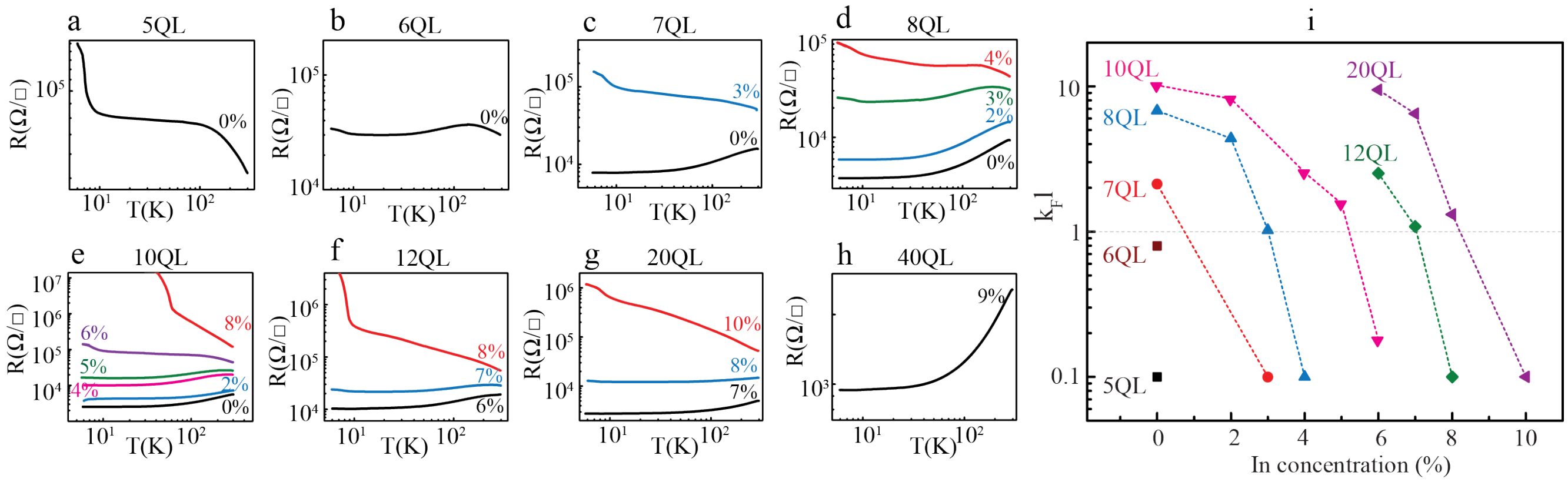

**Figure 1.** Sheet resistance as a function of temperature. (a) 5 QL with 0% In. (b) 6 QL with 0% In. (c) 7 QL with 0% and 3% In. (d) 8 QL with 0%, 2%, 3% and 4% In. (e) 10 QL with 0%, 2%, 4%, 5%, 6% and 8% In. (f) 12 QL with 6%, 7% and 8% In. (g) 20 QL with 7%, 8% and 10% In h, 40 QL with 9% In. (i) $k_F l$ for different thicknesses with different In concentration. The intersection of each curve with the dashed line at $k_F l = 1$ is roughly where the metal-to-insulator transition occurs.

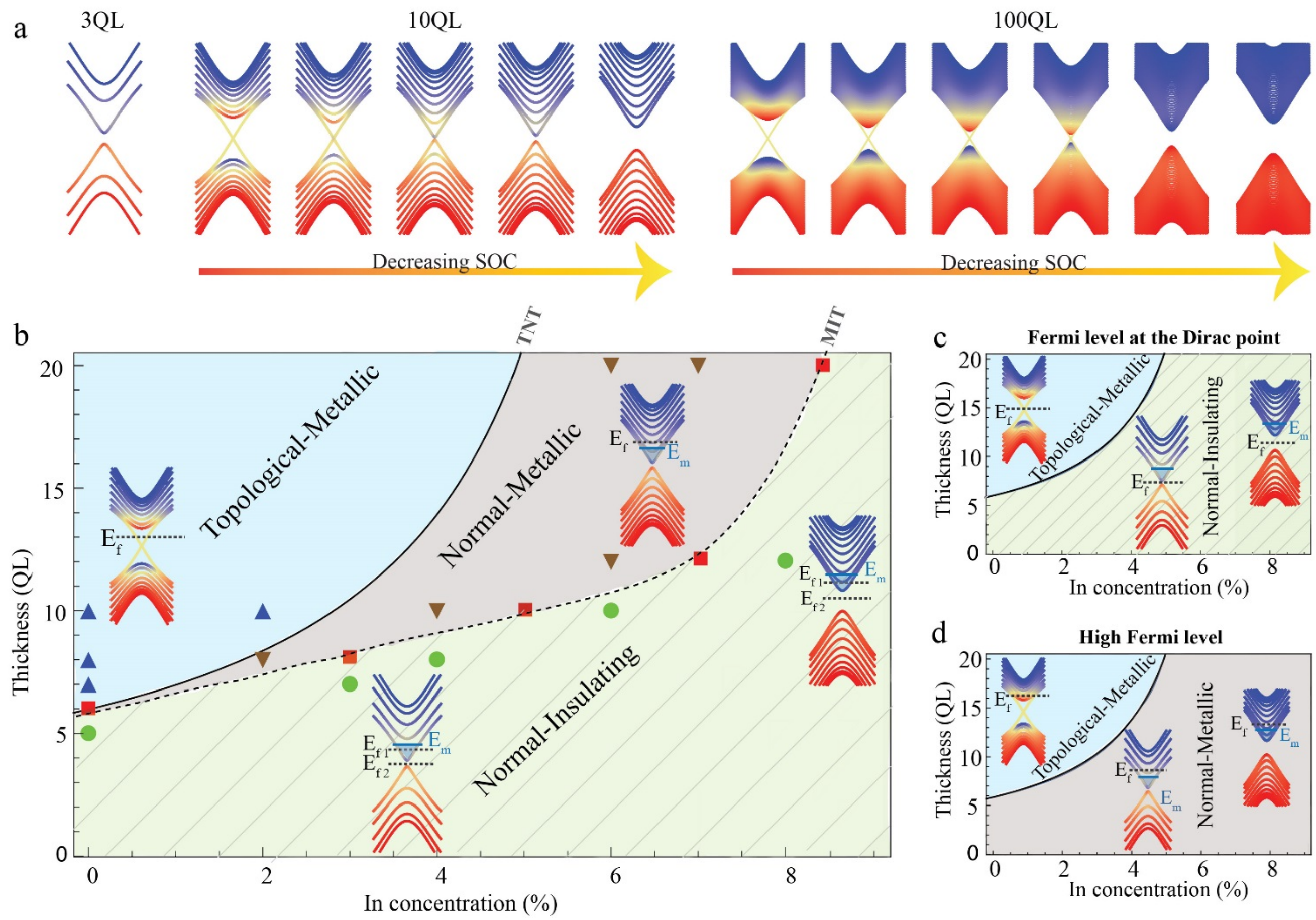

**Figure 2.** Phase transition schematic and phase diagram. (a) Schematic of the TPT process for different thicknesses and as a function of SOC weakening. (b) Phase diagram of the current experiment (the dashed line is a guide to eye for the MIT boundary). Blue triangles correspond to topological-metallic data points where $k_F l > 1$, green circles are normal-insulating data points with $k_F l < 1$, brown triangles show normal-metallic data points with $k_F l > 1$, red squares with $k_F l = 1$ taken from Figure 1i and the MIT line goes through these points. In the normal insulating region, $E_{F1}$ and $E_{F2}$ represent two possible locations of the surface Fermi level, the former below the mobility edge and the latter inside the surface hybridization gap. $E_m$ represents the mobility edge. (c) Phase diagram of the ideal case with the Fermi level at the Dirac point. (d) Phase diagram when the surface Fermi level is high and above the bottom of the conduction band, which corresponds to the previous generation $(Bi_{1-x}In_x)_2Se_3$ samples[14,15].

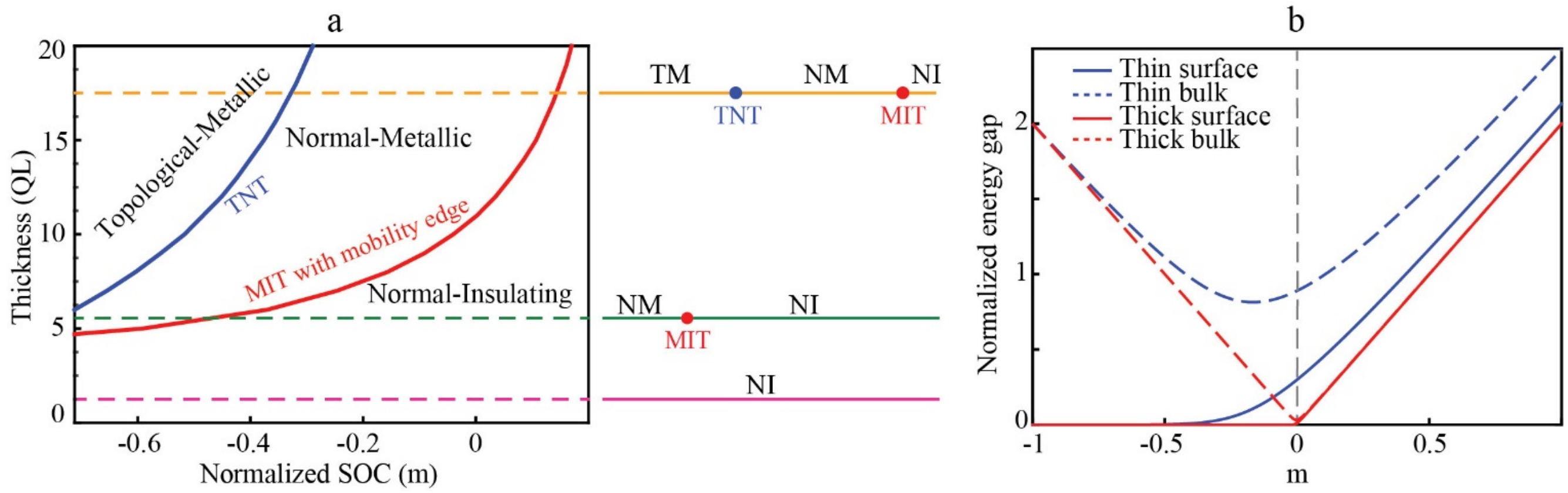

**Figure 3.** Simulated phase diagram and thickness-dependence of the surface and bulk gap developments through the transition. (a) The theoretical phase diagram of TPT in the thin regime. The phase boundaries are determined using the four-band Dirac model (see Supporting Information Section II, Figure S6 for details). The right panel shows the various phases and the transitions between them. (b) The critical point for TNT is well-defined in thick samples, but is not indisputably definable in thin samples. The vertical dashed line (m = 0) shows the phase boundary in the infinite-size limit. Both axes are normalized by the Dirac velocity in this model (Supporting Information Section II for details)

**Supporting Information.**

Experimental details (Section I), numerical simulation (Section II), animations (Section III) and Figures S1-S6.

AUTHOR INFORMATION

**Present Addresses**

‖Current address: Department of Materials Science and Engineering, Pennsylvania State University, University Park, Pennsylvania 16802, USA

**Author contributions**

M.S and S.O conceived the project. M.S synthesized the samples and performed transport measurements and analyzed the data with the assistance of N.K, M.B, and J.M. H.S carried out the theoretical modeling. M.S., H.S., and S.O. wrote the manuscript with inputs from other authors. All authors contributed to the scientific analysis and manuscript revisions.

**Funding Sources**

National Science Foundation, Gordon and Betty Moore Foundation.

ACKNOWLEDGMENT

This work was supported by NSF (DMR-1308142 and EFMA-1542798), and Gordon and Betty Moore Foundation's EPiQS Initiative (GBMF4418).

[§]*These authors contributed equally to this work*

ABBREVIATIONS

TI, topological insulator; TPT, topological phase transition; MIT, metal to insulator transition; TNT, topological to normal transition; TM, topological-metallic, NM, normal-metallic; NI, normal-insulating; SOC, spin orbit coupling.

Table of Contents Graphic and Synopsis:

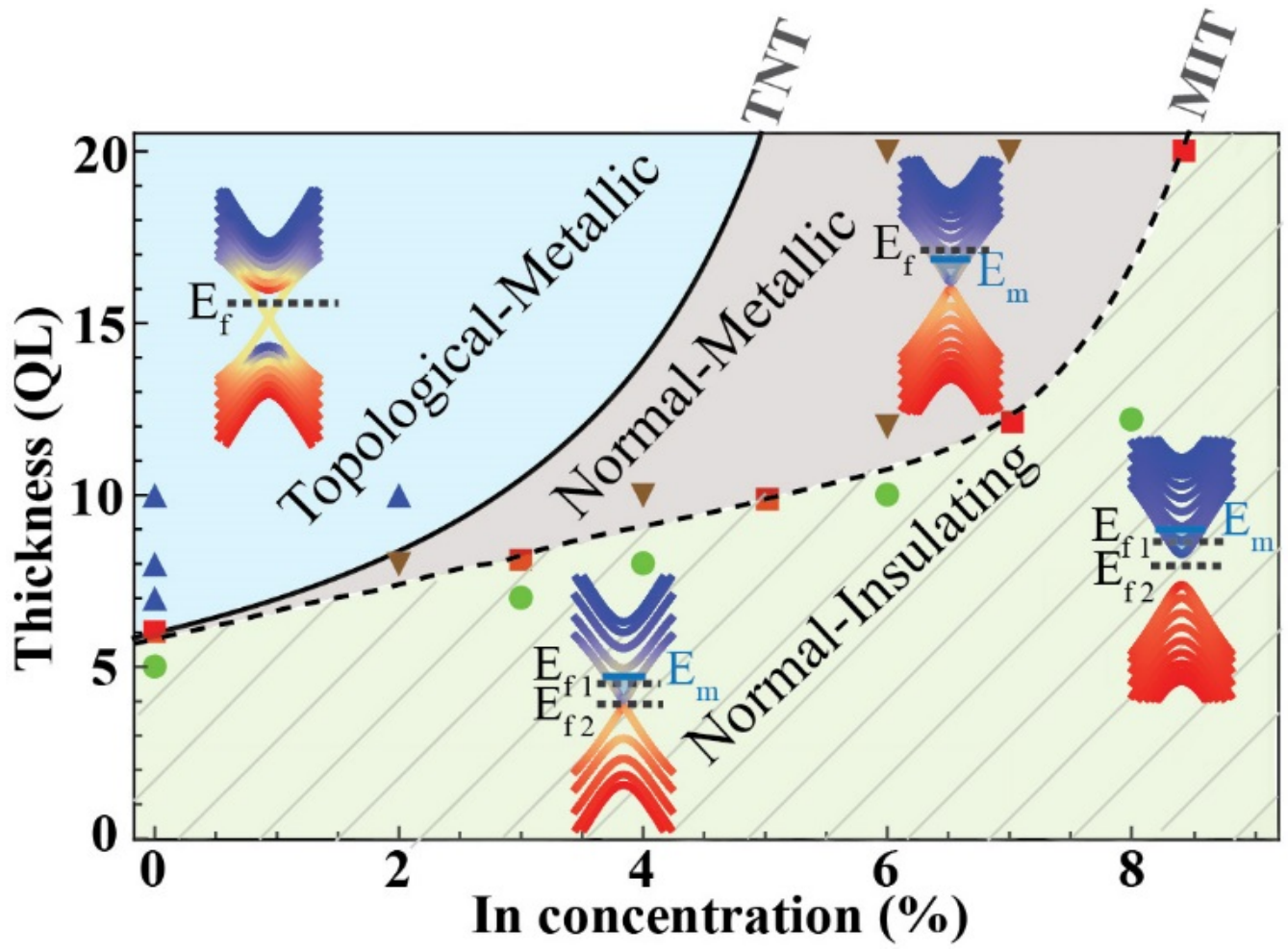

## Supporting Information

# Finite-size and composition-driven topological phase transition in $(\mathbf{Bi}_{1-x}\mathbf{In}_x)_2\mathbf{Se}_3$ thin films

Maryam Salehi§,[1] Hassan Shapourian§,[2] Nikesh Koirala,[3]
Matthew J. Brahlek,[3] Jisoo Moon,[3] and Seongshik Oh[3,*]

[1] *Department of Materials Science and Engineering, Rutgers, The State University of New Jersey, Piscataway New Jersey 08854, U.S.A.*

[2] *Department of Physics, University of Illinois at Urbana-Champaign, Urbana Illinois 61801, U.S.A.*

[3] *Department of Physics and Astronomy, Rutgers, The State University of New Jersey, Piscataway New Jersey 08854, U.S.A.*

§ These authors contributed equally to this work.

* To whom correspondence should be addressed. Email: ohsean@physics.rutgers.edu,
Fax: 732-445-4343

**CONTENTS**

## I. EXPERIMENTAL DETAILS

In this appendix, we present key features of growth and transport properties of the buffer layer-based $(Bi_{1-x}In_x)_2Se_3$ thin films followed by a comparison of sheet resistance of some of these samples with conventionally grown samples.

### A. Growth Methods

We should note that only one of the existing three phases of $In_2Se_3$ shares similar structure with $Bi_2Se_3$; however, unlike $Bi_2Se_3$, $In_2Se_3$ grows in a polymorphic fashion on sapphire. To circumvent this issue, 3QL $Bi_2Se_3$ is deposited at 135 °C which serves as a template for $In_2Se_3$ to be deposited at 300 °C. The whole heterostructure is then heated to 600 °C where the conducting 3QL $Bi_2Se_3$ evaporates and diffuses out of the $In_2Se_3$ layer, leaving behind high quality and insulating buffer layer. This is then followed by deposition of 20 QL of 50% solid solution of $Bi_2Se_3$ and $In_2Se_3$ ($(Bi_{0.5}In_{0.5})_2Se_3$) at 275 °C, which not only works as a template with less lattice mismatch (1.6 % lattice mismatch in comparison to 3.3% for $In_2Se_3$ and 14% for sapphire), but also suppresses the In diffusion into the top layer $(Bi_{1-x}In_x)_2Se_3$ to be deposited at 275 °C. Figure S1(**a**) shows a schematic representation of the growth procedure of $(Bi_{1-x}In_x)_2Se_3$ grown on 20 QL $In_2Se_3$/20 QL $(Bi_{0.5}In_{0.5})_2Se_3$ buffer layers. Finally, amorphous $MoO_3$ capping was deposited in situ and at almost room temperature, serving as an electron-depleting capping layer.

The reflection high-energy electron diffraction (RHEED) in Figure S1(**b**) indicates a two-dimensional highly crystalline growth of $(Bi_{1-x}In_x)_2Se_3$ layer. Furthermore, Figure S1(**c**) shows high-angle annular dark-field scanning transmission electron microscopy (HAADF-STEM) image of a 50 QL $Bi_2Se_3$ film grown on buffer layer, indicating highly ordered growth with sharp interfaces which is mostly due to substantially suppressed interfacial defects. Angle resolved photo-emission spectroscopy (ARPES) in Figure S1(**d**), clearly shows topological surface states (TSS) with the Fermi level $E_F$ locating in the bulk gap and only ~0.17 eV above the Dirac point, confirming that the sample is bulk-insulating topological insulator with only TSS being occupied. With the use of $MoO_3$ the Fermi level can be pushed even further towards the Dirac node. For more details on buffer-based $Bi_2Se_3$ see Ref. (1).

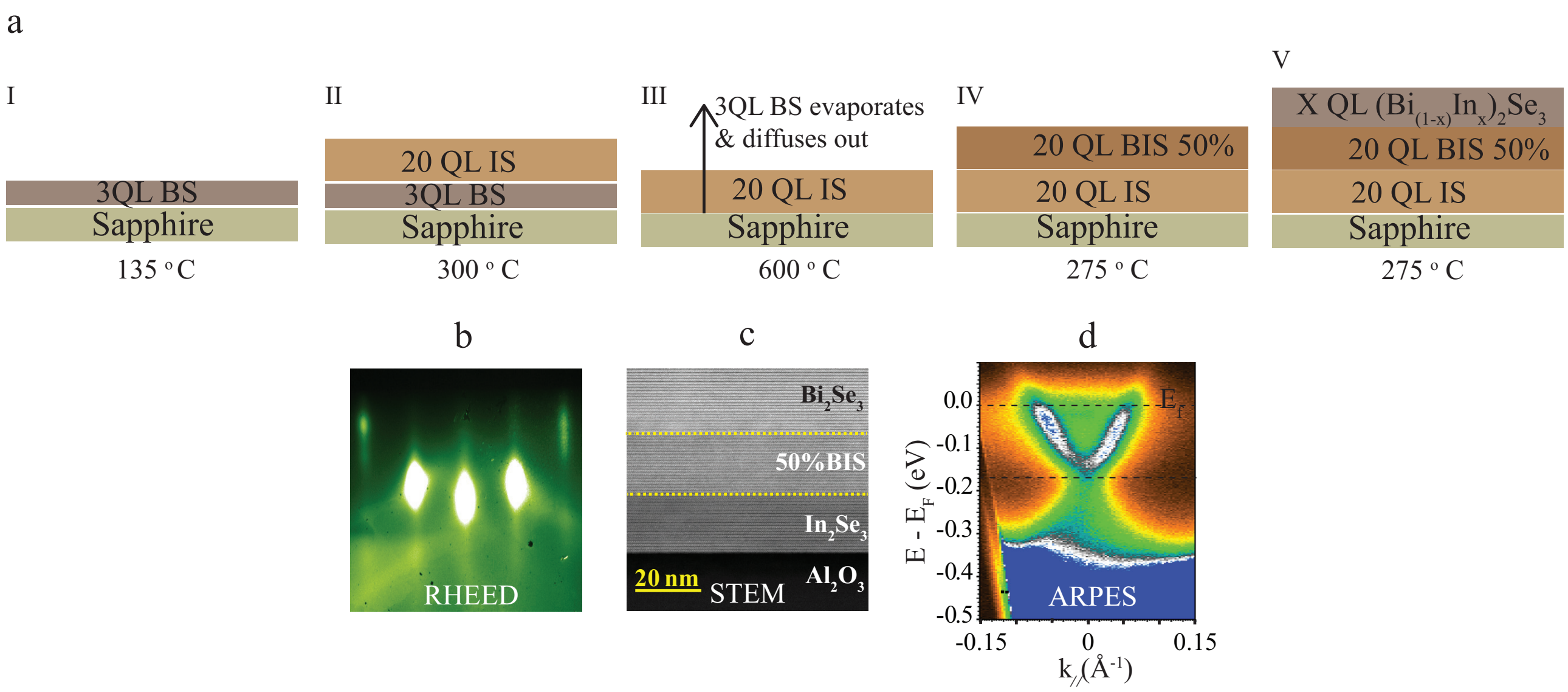

Figure S1. **a,** Schematic of growth process of $(\mathrm{Bi}_{1-x}\mathrm{In}_x)_2\mathrm{Se}_3$ films on the 20 QL $\mathrm{In}_2\mathrm{Se}_3$/20 QL $(\mathrm{Bi}_{0.5}\mathrm{In}_{0.5})_2\mathrm{Se}_3$ buffer layer (BIS-BL). **b,** RHEED of $(\mathrm{Bi}_{1-x}\mathrm{In}_x)_2\mathrm{Se}_3$ shows high quality 2D growth. **c,** HAADF-STEM image shows sharp interfaces. **d,** ARPES shows the surface states and the location of $E_F$ in the bulk gap.

## B. Transport

Figure S2 summarizes sheet carrier density ($n_{\mathrm{sheet}}$) in the top panel and mobility ($\mu$) in the bottom panel for samples with different thicknesses and In concentrations. For insulating samples, they are measured at high enough temperatures before the carriers are immeasurably frozen. Since the surface $E_F$ of these samples is in the bulk gap, we can calculate the 2D surface $E_F$ by considering only the surface band, which can be approximated up to quadratic order by $E_F = Ak_F + Bk_F^2$ ($A = 2.02$ eVÅ and $B = 10.44$ eVÅ$^2$), where 2D Fermi vector $k_F = (2\pi n_{\mathrm{sheet}})^{1/2}$ (assuming similar carrier density for two TSSs on the top and bottom surfaces) (2, 3). Based on these equations, $E_F$ is estimated to be between 45 to 70 meV above the Dirac point for the entire thickness range 6QL to 40QL in the absence of In substitution (for reference, the bottom of the conduction band is 220 meV above the Dirac point).

In addition, the metal-insulator transition (MIT) can be also identified through the weak anti-localization (WAL) analysis, which appears as a cusp in magneto-conductance measurement at low magnetic field. According to the Hikami-Larkin-Nagaoka (HLN), the WAL effect in the magneto-conductance data can be fitted by $\Delta G(B) = -\tilde{A}e^2/(2\pi h)[\ln(B_\phi/B) - \Psi(1/2 + B_\phi/B)]$, where $h$ is the Planck constant and $\Psi$(x) is the digamma function, using the two fitting parameters, $\tilde{A}$ and $B_\phi$ (4). $\tilde{A}$ represents the number of conducting 2D channels ($\tilde{A}$= 1 for each channel) and thus, relatively insensitive to material details, whereas $B_\phi$, the de-phasing field, varies substantially depending on the level of material qualities. Accordingly, as the film goes through MIT, we expect the WAL channel ($\tilde{A}$) to collapse to nearly zero, which is also confirmed in magneto-conductance data of a 10 QL-thick sample (Figure S2(**b**)). The WAL channel dependence on sample thickness and In concentration in Figure S2(**c**) provides a consistent and complementary view on the MIT as analyzed above using other methods.

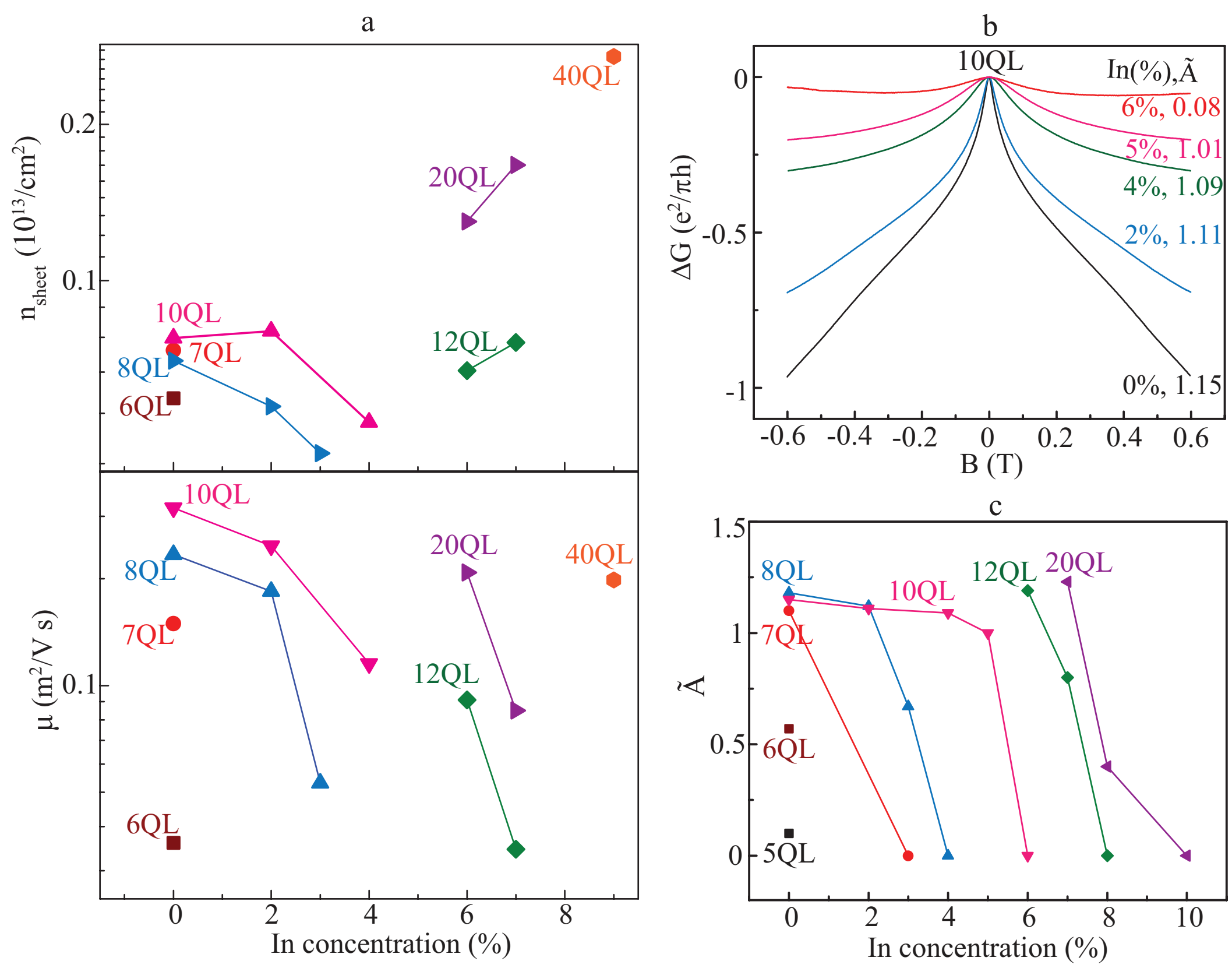

Figure S2. **a**, $n_{\mathrm{sheet}}$ and $\mu$ for different thicknesses as a function of In concentration. Longitudinal resistance vs. temperature of $Bi_2Se_3$ grown on buffer layer for different In concentration. **b**, Magneto-conductance for 10QL sample with different In concentration. **c**, for different thicknesses with different In concentration. heet.

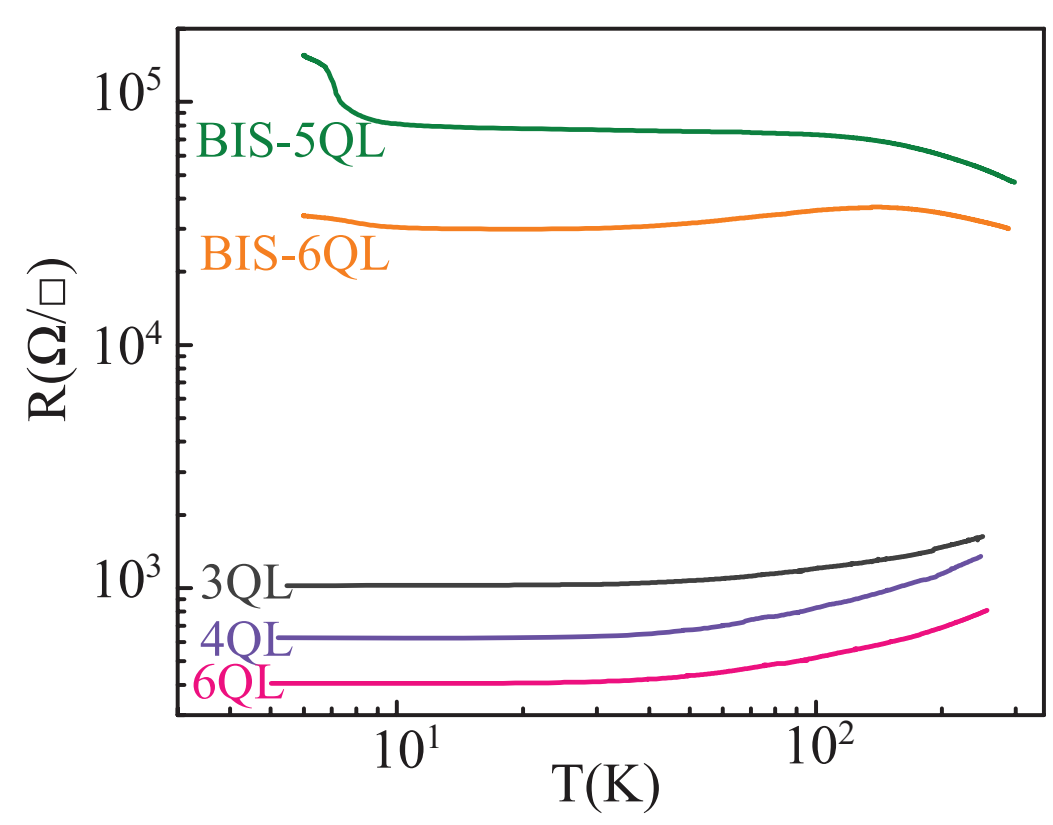

Figure S3. Comparison of sheet resistance as a function temperature for buffer layer-based vs. conventionally grown thin films.The BIS refers to $In_2Se_3/(Bi_{0.5}In_{0.5})_2Se_3$ buffer layer.

Moreover, Figure S3 shows the comparison of sheet resistance of 3, 4, and 6QL thick regularly grown $Bi_2Se_3$ on sapphire substrate (5) with 5 and 6QL thick $Bi_2Se_3$ grown on 20QL $In_2Se_3$/20QL $(Bi_{0.5}In_{0.5})_2Se_3$ buffer layer and capped by $MoO_3$ , indicating that the buffer layer-based samples are more resistive due to their surface Fermi level being much closer to the Dirac point.

## II. NUMERICAL SIMULATION OF THE TOPOLOGICAL PHASE TRANSITION

In this appendix, we theoretically study the finite size effect on gap opening and the topological phase transition in $(Bi_{1-x}In_x)_2Se_3$ thin films. We use the four-band Dirac model to simulate the low energy properties of this material. This model was introduced in (6) and has been used to study interesting electromagnetic properties of topological insulators such as the quantized anomalous Hall effect, the magnetoelectric (axion) effect, and the Faraday effect (7, 8). Near $\Gamma$ point, the effective low energy Hamiltonian is written as

$$H(\mathbf{k}) \approx \epsilon_0(\mathbf{k})\mathbf{I}_{4\times4} + \begin{pmatrix} \mathcal{M}(\mathbf{k}) & A_1k_z & 0 & A_2k_- \\ A_1k_z & -\mathcal{M}(\mathbf{k}) & A_2k_- & 0 \\ 0 & A_2k_+ & \mathcal{M}(\mathbf{k}) & -A_1k_z \\ A_2k_+ & 0 & -A_1k_z & -\mathcal{M}(\mathbf{k}) \end{pmatrix} \tag{1}$$

in the basis $(|P1_z^+,\uparrow\rangle, |P2_z^-,\uparrow\rangle, |P1_z^+,\downarrow\rangle, |P2_z^-,\downarrow\rangle)$ where $k_\pm = k_x \pm ik_y$, $\epsilon_0(\mathbf{k}) = C + D_1k_z^2 + D_2k_\perp^2$ and $\mathcal{M}(k) = m - B_1k_z^2 - B_2k_\perp^2$. Note that $P1_z^+$ and $P2_z^-$ represent the $p_z$ orbitals of bismuth and selenium with positive (+1) and negative (−1) parity, respectively and $\uparrow$ / $\downarrow$ denotes the electron spin. The parameters $A_i, B_i, C$, and $D_i$ can be chosen carefully to reproduce the band structure near the $\Gamma$ point of the $Bi_2Se_3$ (6). Here, our goal is to illustrate the topological phase transition from $Bi_2Se_3$ to $In_2Se_3$ by tuning the "mass" parameter $m$ which represents the *atomic* spin-orbit coupling strength and is responsible for the band inversion. Note that $A_i$ and $B_i$ terms are physically related to the *inter-atomic* spin-orbit coupling between unlike orbitals and the spin-independent electron hopping between identical orbitals, respectively. It is worth mentioning that the spin-momentum locking is indeed due to $A_i$ terms.

Here, we focus on the qualitative features of the phase transition at finite thickness rather than fine-tuning the parameters to reproduce the exact band structures. One reason is that the physics near the transition point is governed by the low energy theory and must be universal. Along this direction, we make another simplification by neglecting $\epsilon_0(\mathbf{k})$ as it only shifts the entire band structure and has no effect on the phase transition. It is more convenient to write the Hamiltonian Eq. (1) in the standard form of 3D massive Dirac model as

$$H(\mathbf{k}) = A_2(k_x\alpha_x + k_y\alpha_y) + A_1k_z\alpha_z + \mathcal{M}(\mathbf{k})\beta \tag{2}$$

where the Dirac matrices are given by

$$\alpha_s = \tau_1 \otimes \sigma_s = \begin{pmatrix} 0 & \sigma_s \\ \sigma_s & 0 \end{pmatrix}, \qquad \beta = \tau_3 \otimes \mathrm{I} = \begin{pmatrix} \mathrm{I} & 0 \\ 0 & -\mathrm{I} \end{pmatrix}, \qquad \gamma_5 = \tau_1 \otimes \mathrm{I} = \begin{pmatrix} 0 & \mathrm{I} \\ \mathrm{I} & 0 \end{pmatrix},$$

and $s = x, y, z$. In this convention the $\sigma_{x,y,z}$ and $\tau_{1,2,3}$ matrices act on the spin and orbital degrees of freedom respectively. There are two important symmetries present in this model: (i) time-reversal symmetry (TRS) with the operator $\mathcal{T} = i\sigma_2\mathcal{K}$ such that $\sigma_2H(\mathbf{k})\sigma_2 = H^*(-\mathbf{k})$; (ii) inversion symmetry (IS), represented by $\mathcal{I} = \tau_3\mathcal{P}$, such that $\tau_3H(\mathbf{k})\tau_3 = H(-\mathbf{k})$. As the mass parameter $m$ is varied, the Hamiltonian demonstrates a phase transition from the topological phase ($m < 0$) with an inverted band to the non-topological (trivial) phase ($m > 0$) where bands are adiabatically connected to the atomic limit.

A thin film is modeled by imposing an open boundary condition along the $z$ direction. This boundary condition breaks the translational symmetry and hence, $k_z$ is no longer a good quantum number and must be replaced by the gradient operator $-i\partial_z$ in the real-space. To numerically simulate this geometry, we consider a one-dimensional lattice along the $z$ direction where each site represents one quintuple (QL) of $Bi_2Se_3$. In this process, the $\partial_z$ operator must be replaced by a hopping operator along the lattice. Therefore, the Hamiltonian can be recast in the following form

$$H = \begin{pmatrix} h_\perp(\mathbf{k}) & \Delta_z & & & \\ \Delta_z^\dagger & h_\perp(\mathbf{k}) & \Delta_z & & \\ & \Delta_z^\dagger & \ddots & \ddots & \\ & & \ddots & \ddots & \Delta_z \\ & & & \Delta_z^\dagger & h_\perp(\mathbf{k}) \end{pmatrix}$$

for which the matrix row/columns are associated with QLs and only non-zero entries are shown. The inter-plane hopping is $\Delta_z = -iA_1\alpha_z/2 - B_1\beta$, and the in-plane Hamiltonian for each QL is $h_\perp(\mathbf{k}) = A_2(k_x\alpha_x + k_y\alpha_y) + (m + 2B_1 - B_2k_\perp^2)\beta$. We also made further simplification by choosing $A_1 = A_2 = v$, $B_1 = B_2 = B$. Throughout this paper, the energy scale is always mentioned in units of the Dirac velocity $v$ (more precisely $\hbar v \left(\frac{\pi}{a}\right)^2$ where $a$ is the lattice constant). Figure S4 shows the band structure in $k_x = 0$ plane where we use different color schemes to explicitly show the orbital/spin content of the in-gap surface modes as well as the bulk states. In Figure S4(**a**) the energy bands are colored due to their orbital contents in which the pure Se and Bi orbitals are red and blue respectively and fully mixed (50-50) states are shown in yellow. Using the basis introduced in Eq. (2), this color code can be written as $\langle \tau_3 \rangle$, i.e. the expectation value of $\tau_3$ for a given eigenstate. In Figure S4(**b**), the spin texture is illustrated where blue and orange show the opposite spin directions and eqaul superposition is black. We should note that the spin texture cannot be given by $\langle \sigma_{x,y,z} \rangle$ due to the inter-atomic spin-orbit coupling and we must consider a combined operator in spin-orbital basis such as $\tau_1 \sigma_{x,y}$. As we explicitly derive below, $\langle \sigma_{x,y,z} \rangle = 0$ for all surface states while $\langle \tau_1 \sigma_{x,y} \rangle$ is non-zero and shows the spin precession in surface states. In particular, we use $\langle \tau_1 \sigma_y \rangle$ for the color scheme in Figure S4(**b**). In Figure S4(**c**), we distinguish between the bulk and surface states by computing the inverse participation ratio $\text{IP}(\Psi) = \sum_z |\Psi(z)|^4 / \sum_z |\Psi(z)|^2$. It is easy to see that the IP($\Psi$) is maximized (minimized) for confined surface (deconfined bulk) states in the $z$ direction. In this figure, the two extreme colors are black (IP$= 0$) and red (IP$= 1$).

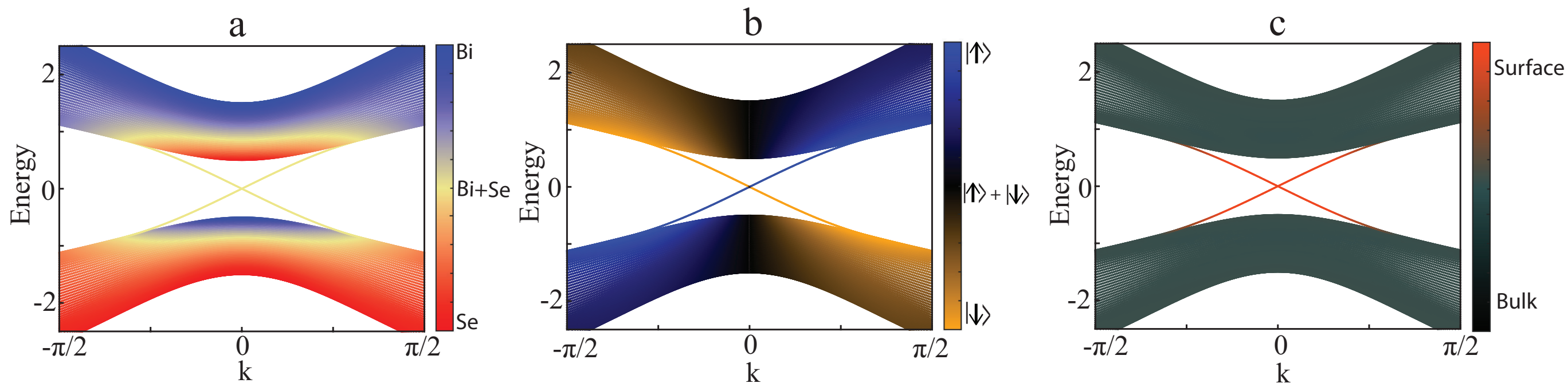

Figure S4. The low energy band structure of a thick TI sample. The color schemes are associated with **a**, orbital contents, **b**, surface spin component and **c**, confinement near the surface boundaries. The thickness is 100 QLs and $m = -0.5$.

As we have seen in Figure 3(b) of the main text, in the thin regime the hybridization of top and bottom surface states causes the gap to open at $m < 0$ which is before the critical boundary in the infinite-size limit at $m = 0$. It is also worth noting that the bulk gap in the thin regime never entirely closes and is only minimized near (slightly after) the surface gap opening point. This is in contrast to the usual expectation that the surface states are protected until the bulk gap closes and reopens again, which is only the case in the thick regime (red curves in Figure 3(b) of the main text). To pinpoint the transition in the finite-size system, we look for a quantity (an order parameter) which reliably measures the band inversion along the topological phase transition. One possibility is to compute the population imbalance in the orbital space and to identify where the band de-inversion occurs. In the basis defined in Eq. (2), this means to compute the expectation value of the pseudo-spin operator $\tau_3$ which returns the difference of probability amplitudes of Bi and Se orbitals for a given eigenstate. The general trend in Figure S5 shows that overall the eigenstates have equal contribution from Bi and Se orbitals deep in the topological phase ($m = -1$) where the TI is the strongest (largest bulk gap) and as we go toward the trivial phase (atomic limit) the eigenstates are populated by Se orbitals, where $\langle\tau_3\rangle \to -1$. Furthermore, the point at which the surface gap begins to open coincides with the inflection point of the $\langle\tau_3\rangle$ that is manifest as a peak in $\frac{d\langle\tau_3\rangle}{dm}$ (Figure S5(**c**)). Interestingly, the peak gets broader as the TI becomes thinner and the derivative diverges at $m = 0$ in the infinite-size limit. Another way to interpret this important observation is that the critical point in the infinite-size limit turns into a gradual crossover in the thin regime. We should note that this crossover in the thin film occurs before the infinite-size critical point and $m = 0$ is not the critical point anymore. Aside from these facts, it is worth noting that in our simulations, the system becomes normal once the surface gap opens and we do not observe a topological bulk with hybridized surface states. This observation is justified as there is no bulk gap closing anywhere at larger values of $m$ beyond the hybridization point.

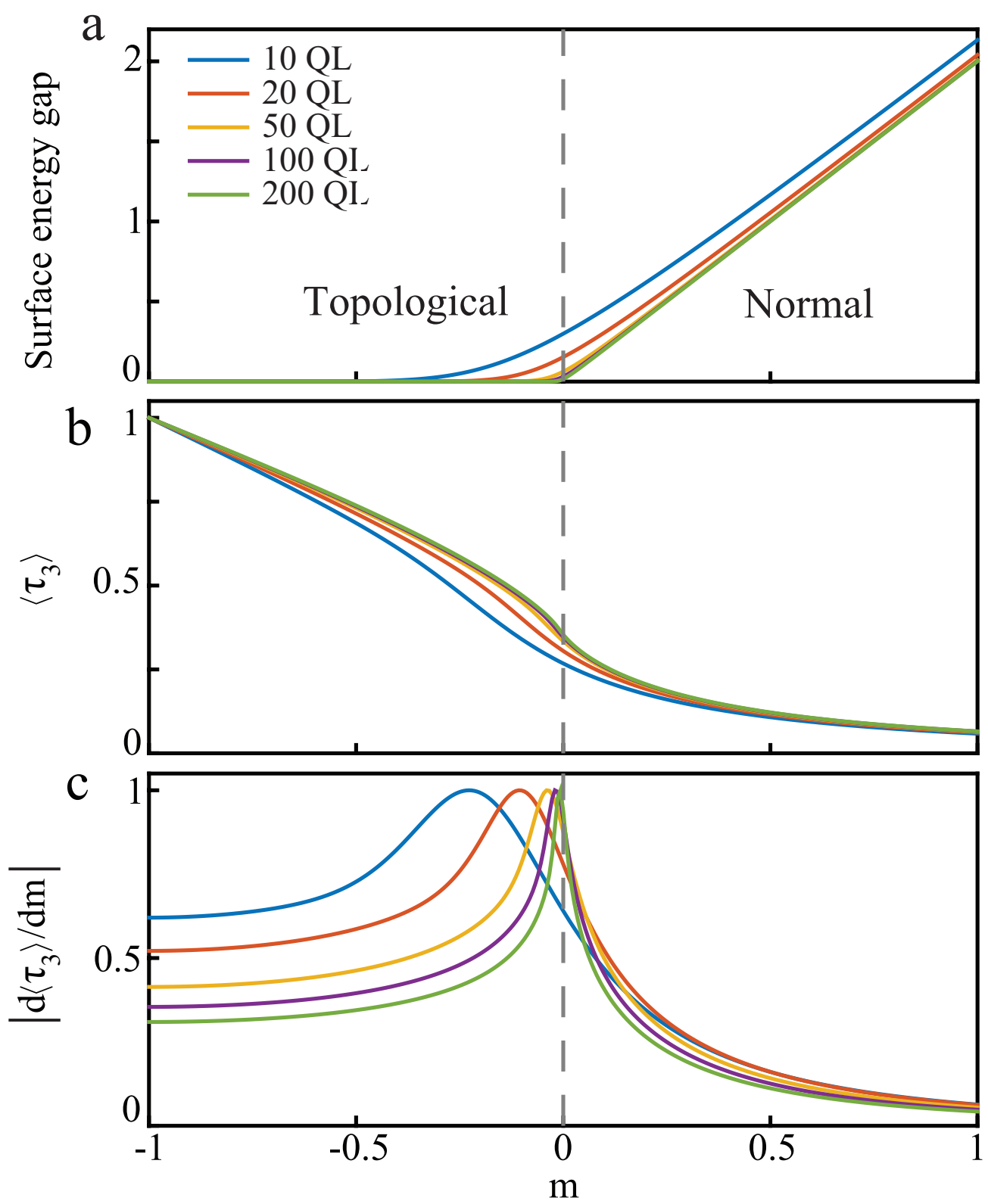

Figure S5. The surface gap opening and the band inversion as a function of $m$ for different thicknesses. **a**, the gap in the surface states spectrum, **b**, the expectation value of the band inversion operator $\langle\tau_3\rangle$, the vertical axis is shifted up by 1. **c**, the normalized derivative $|\frac{d\langle\tau_3\rangle}{dm}|$.

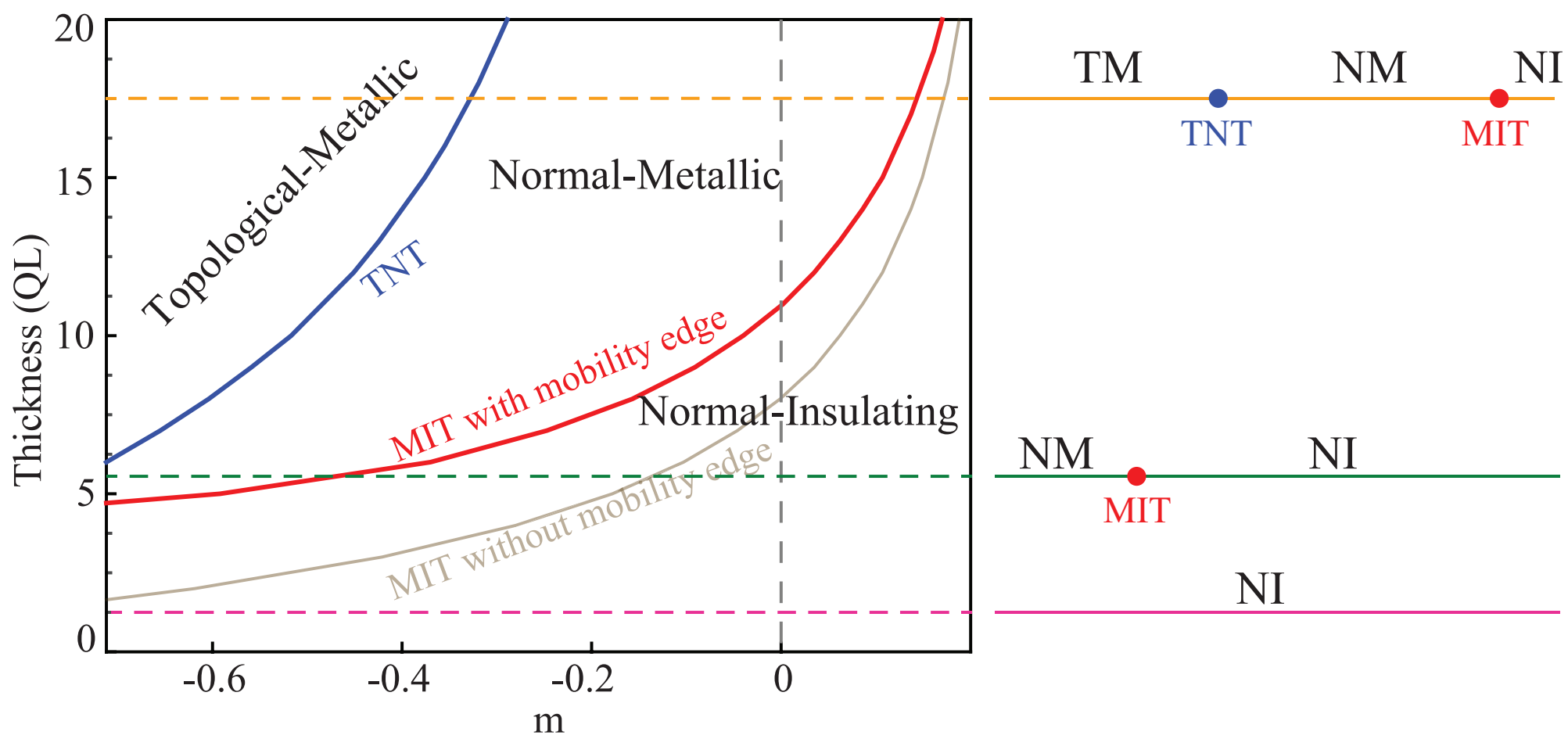

Figure S6. The phase diagram of TI in the thin regime. The phase boundaries are determined using the four-band Dirac model where a metallic (insulating) phase is identified by having non-zero (zero) density of states at the Fermi level. The right panel shows the various phases (and the transitions between them) encountered as one move along a horizontal line in the phase diagram. The Fermi levels are set at $E_F = 0$ for the TNT boundary and $E_F = 0.5$ for the MIT without mobility edge boundary. For MIT with mobility edge boundary, we set $E_0 = 0.9$ and $\lambda = 8$QL.

We also examine the four-band model to see if it reproduces the suggested experimental phase-diagram. Interestingly, a consistent phase diagram can be derived for any generic set of parameters as shown in Figure S6. To map out the phase boundaries, we numerically compute the full spectrum of a finite-size sample and check whether the density of states (DOS) at the Fermi level is zero or not. The zero or non-zero DOS correspond to the insulating or metallic phases respectively. The boundary for the topological to normal insulator transition (TNT) is found by fixing the Fermi level at zero and the boundary for the metal to insulator transition (MIT) is found by fixing the Fermi level at a non-zero value. Inspired by the experimental findings, we use a simple phenomenological model to simulate the effect of disorder by introducing a mobility edge. The mobility edge is defined as the boundary between the extended (conducting) and localized (insulating) states in the conduction band. The mobility edge smoothly moves down towards the bottom of the conduction band as the sample becomes thicker likely due to reduced inter-surface scattering effect. For our simulations, the distance between the bottom of the conduction band and the mobility edge is chosen to be exponentially varied $\delta E_{edge} = E_0 \exp(-t/\lambda)$ as a function of the sample thickness $t$, where $E_0$ (the maximum mobility edge region) and $\lambda$ (the length scale to interpolate the mobility edge behavior) are two phenomenological parameters introduced to model the gradual suppression of the mobility edge region as the sample is made thicker.

### A. Derivation of Dirac Surface Modes

In the remainder of this section, we derive the surface state wave functions and show that each surface Dirac spectrum can be written in an effective two-dimensional Hilbert space of a Kramer's doublet. To find the spin/orbital content of the surface states we need to solve the domain-wall problem where a boundary separates the topological phase from the trivial one. This is modeled by a spatial-varying mass term $m(z)$. We shall solve for the in-gap states near the TI-nonTI interface at $z = 0$ where the upper-half ($z > 0$) is filled with TI $m(z) < 0$ and the lower-half ($z < 0$) is filled with non-TI $m(z) > 0$. A similar solution can be found for the other boundary surface. Let us start by finding the zero-mode states on the TI side. The Schröedinger equation reads

$$(-i\tau_1\sigma_z v\partial_z + m\tau_3)\,|\Psi\rangle = 0 \tag{3}$$

the solution of which can be written as $|\Psi\rangle = |\psi_p\rangle \otimes |\psi_s\rangle$ where $|\psi_p\rangle$ is in the orbital space and $|\psi_s\rangle$ is in the spin space. The following equation

$$\begin{pmatrix} m & \mp iv\partial_z \\ \mp iv\partial_z & -m \end{pmatrix} |\psi_p\rangle = 0 \tag{4}$$

has two solutions

$$|\Psi_1\rangle = \frac{e^{-mz/v}}{\sqrt{2}} \begin{pmatrix} 1 \\ i \end{pmatrix} \otimes |\uparrow\rangle, \tag{5}$$

$$|\Psi_2\rangle = \frac{e^{-mz/v}}{\sqrt{2}} \begin{pmatrix} 1 \\ -i \end{pmatrix} \otimes |\downarrow\rangle, \tag{6}$$

Note that these two wave-functions form a Kramer's doublet, i.e. one is the time-reversal conjugate of the other. Next, we are going to derive the full spectrum of the Dirac surface states. To this end, we look for in-gap state solutions of non-zero in-plane momenta $\mathbf{k} = (k_x, k_y)$,

$$(\tau_1\sigma_x vk_x + \tau_1\sigma_y vk_y - i\tau_1\sigma_z v\partial_z + m\tau_3)|\Psi\rangle = \epsilon_{\mathbf{k}}|\Psi_{\mathbf{k}}\rangle. \tag{7}$$

To find the near-zero energy states of the above equation, we rewrite the Hamiltonian in the Kramer's doublet space. Note that the application of last two operators inside the above parenthesis on these Kramer's doublet is already zero. Equivalently, one can write the following ansatz as the solution

$$|\Psi_{\mathbf{k}}\rangle = \alpha_{\mathbf{k}}|\Psi_1\rangle + \beta_{\mathbf{k}}|\Psi_2\rangle \tag{8}$$

where the Schrödinger equation in the Kramer's doublet basis becomes

$$\begin{pmatrix} 0 & -v(ik_x+k_y) \\ -v(-ik_x+k_y) & 0 \end{pmatrix} \begin{pmatrix} \alpha_{\mathbf{k}} \\ \beta_{\mathbf{k}} \end{pmatrix} = \epsilon_{\mathbf{k}} \begin{pmatrix} \alpha_{\mathbf{k}} \\ \beta_{\mathbf{k}} \end{pmatrix}. \tag{9}$$

Thus, the surface eigenstates are found to be

$$|\Psi_{\mathbf{k},\pm}\rangle = \frac{1}{\sqrt{2}}(|\Psi_1\rangle \pm ie^{i\theta_{\mathbf{k}}}|\Psi_2\rangle) \tag{10}$$

where $\tan\theta_{\mathbf{k}} = k_y/k_x$ and the dispersion is linear $\epsilon_{\mathbf{k}} = \pm v|\mathbf{k}|$. It is easy to see that the expectation value of "physical" spin operators $\sigma_{x,y,z}$ for the surface states are all zero. In order to unravel the spin texture we need to project onto the Kramer's doublet subspace by computing the operators $O_1 = \tau_1\sigma_x$ or $O_2 = \tau_1\sigma_y$. Therefore, we have

$$\langle\Psi_{\mathbf{k},s}|O_1|\Psi_{\mathbf{k},s}\rangle = is(\alpha_{\mathbf{k}}\beta^*_{\mathbf{k}} - \alpha^*_{\mathbf{k}}\beta_{\mathbf{k}}) = s\cos\theta_k \tag{11}$$

$$\langle\Psi_{\mathbf{k},s}|O_2|\Psi_{\mathbf{k},s}\rangle = s(\alpha_{\mathbf{k}}\beta^*_{\mathbf{k}} + \alpha^*_{\mathbf{k}}\beta_{\mathbf{k}}) = s\sin\theta_k \tag{12}$$

where $s = \pm$. This explicitly shows the spin precession as one make a round trip over the Dirac surface spectrum (see also Figure S4(**b**)).

## III. ANIMATIONS

Using the color scheme proposed in Sec. II and shown in Figure S4, we provide six animations to illustrate the band evolution across the topological phase transition as the atomic spin-orbit coupling strength $m$ is tuned for a thin and a thick sample. The tuning direction is from the topological phase where $m < 0$ (i.e. $Bi_2Se_3$) to the trivial phase where $m > 0$ (i.e. $In_2Se_3$); this is to model the process of adding In to the $Bi_2Se_3$ crystal. For each thin and thick sample, the three animations are associated with three different color schemes (orbital content, spin texture and confinement near the boundary surfaces) to explicitly show how each property is evolved as a function of the spin-orbit coupling. To our knowledge, this is the most insightful and pedagogical way of understanding and representing the behavior of TIs near the TPT point. In the following, we briefly explain the main features in each file.

1. **Orbital content as color**

   (a) `orbital_thick.mp4`: It initially shows the band inversion near the $\Gamma$ point where the surface states appear in the gap. Note that the surface states are colored as equal contribution of Bi and Se p-orbitals. As $m$ is increased, the bulk gap gets smaller until the conduction and valence bands meet at the critical point and eventually they turn into their original band color by passing through yellow (mixture of odd and even parity). Notice that the surface states survive all the way to the critical point when the bulk gap closes.

   (b) `orbital_thin.mp4`: The important difference here is that the bulk gap never closes and the surface Dirac node opens before $m = 0$.

2. **Confinement as color**

   (a) `local_thick.mp4`: It shows that the surface states are only confined near the boundaries and they turn grey (deconfined) gradually which means that the width of surface wave-functions grow gradually. Note that the surface states do not open a gap right after they become grey (i.e. their width is large but relatively small compared to the sample thickness).

   (b) `local_thin.mp4`: The same trend is observed here except for the fact that the surface states start to open a gap at $\Gamma$ point while the rest of the surface band is not even completely grey. This is because once the width of surface modes becomes comparable with the sample thickness they hybridize and the gap opens.

3. **Spin component as color**

   (a) `spin_thick.mp4`: The spin texture in the surface states is evident. Note that the spin-momentum locking is due to the inter-atomic spin-orbit coupling which is not tuned in our simulations. It is worth noting that the emergence of surface states is due to strong atomic spin-orbit coupling and tuning this parameter is responsible for the TPT. Therefore, we do not expect to see a significant difference in the spin texture of the bulk states in either topological or trivial phases.

   (b) `spin_thin.mp4`: Besides the fact that the gap opening occurs before the bulk transition point $m = 0$, there is no major difference between thin and thick regimes.